\documentstyle[12pt,graphicx]{article}

\begin{document}

\begin{center}
{\LARGE Soft and Hard Pomeron in the Structure Function of the Proton
at Low $x$ and Low $Q^2$\\}
\vskip 20pt
{\Large
 U. D'Alesio, A. Metz and H.J. Pirner}\\
\bigskip 
{\it
Institut f\"ur Theoretische Physik der Universit\"at Heidelberg,\\
Philosophenweg 19, 69120 Heidelberg, Germany\\}
\bigskip
\end{center}

\begin{abstract}
We study inclusive electroproduction on the proton at low $x$ and
low $Q^2$ using a soft and a hard Pomeron.
The contribution of the soft Pomeron is based on the Stochastic 
Vacuum Model, in which a nonperturbative dipole-dipole cross 
section can be calculated by means of a gauge invariant gluon 
field strength correlator.
To model the hard Pomeron exchange we phenomenologically extend
the leading order evolution of a power-behaved structure function, 
$F_2 \propto x^{- \lambda}$, proposed by L\'opez and Yndur\'ain.
This extension allows to consider both the case $Q^2 = 0$ and the
region of higher $Q^2$ on the basis of the same parametrization.
A good simultaneous fit to the data on $F_2$ and on the cross section 
$\sigma_{\gamma p}$ of real photoproduction is obtained for 
$\lambda=0.37$.
With four parameters we achieve a $\chi^2/\textrm{d.o.f.} = 0.98$ for 
222 data points.
In addition, we use our model of the inclusive $\gamma^{\ast} p$ 
interaction to compute the longitudinal structure function $F_L$.
\end{abstract}
\bigskip

\section{Introduction}
Since the start of HERA exciting new information on the proton structure 
at very small $x$ has been produced. 
Recently, the low $x$ study has been extended to cover very small 
virtualities $Q^2$ \cite{h1a_97,zeusa_97}. 
This kinematical region is of particular interest because a transition 
from the purely perturbative scaling violations at large $Q^2$ to 
different physics at small $Q^2$ can be observed.

We consider the photon-proton collision in the {\it cm} frame.
In this frame, the photon acquires a structure leading to the  
interaction of two structured objects.
Accordingly, at least at high {\it cm} energy and low $Q^2$, 
the $\gamma^{\ast} p$ interaction has strong similarities to the 
hadron-hadron interaction.
Our consideration of the soft Pomeron is based upon the fluctuation 
of the photon into a $q \bar{q}$ dipole.
Higher Fock states (e.g. $q \bar{q}g$) in the wave function of the
photon may lead to a hard Pomeron behaviour of the $\gamma^{\ast}p$
cross section.

In the framework of a model containing a soft and a hard Pomeron we 
perform in the present paper a fit to HERA data on $F_2$ together with 
data from NMC and E665 in a kinematical window given by
$0.11 \, \textrm{GeV}^2 \le Q^2 \le 6.5 \, \textrm{GeV}^2$, 
$x \le 0.01$, and $W \ge 10 \, \textrm{GeV}$, with $W$ representing the 
$\gamma^{\ast} p$ {\it cm} energy.
Moreover, all data on the total absorption cross section 
$\sigma_{\gamma p}$ of real photons with $W \ge 10 \, \textrm{GeV}$ 
are considered. 
The fit includes four free parameters.
Our investigation is a natural extension of previous 
work~\cite{gousset_98}, where $F_2$ at fixed $W$ 
$(20 \, \textrm{GeV})$ has been studied as a function of $Q^2$.
In Ref.~\cite{gousset_98} only a soft Pomeron contribution, derived in
the Stochastic Vacuum Model (SVM)~\cite{dosch_87,dosch_88}, has been 
taken into account.
The aim of the present paper is to study the importance of the soft 
Pomeron as given by the SVM in the low $x$ and low $Q^2$ region of
HERA.

The SVM is a specific model of nonperturbative QCD. 
It lives on the assumption that the infrared behaviour of QCD 
can be approximated by a Gaussian stochastic process.
As central quantity of the SVM serves the correlator of the gluon field 
strength (nonlocal gluon condensate), which consists of an Abelian and
a non-Abelian part, where the non-Abelian correlator gives rise to
linear confinement in terms of the Wilson area law.
Three parameters determine the correlator:
the overall normalization is given by the local gluon condensate 
$\langle g^2 FF \rangle$, while the correlation length $a$ fixes
the shape of the correlator in coordinate space.
Finally, the parameter $\kappa$ regulates the relative weigth of the 
Abelian and the non-Abelian term.
These parameters are proper quantities of nonperturbative QCD and can 
be obtained from lattice simulations.
More details on the technical aspects of the SVM may be found e.g. in 
Ref.~\cite{dosch_94a}.

By means of the eikonal approximation an expression for the 
dipole-dipole scattering amplitude of the SVM has been 
derived~\cite{dosch_94b}. 
Any diffractive reaction involving hadrons or photons can be calculated
by means of the dipole-dipole amplitude and the wave functions of the 
particles.
Besides studies on $F_2$~\cite{gousset_98,rueter_98b}, the SVM has 
been applied to the hadron-hadron 
scattering~\cite{dosch_94b,rueter_96,berger_98},
photo- and electroproduction of vector 
mesons~\cite{dosch_97,kulzinger_98} and of $\pi^0$~\cite{rueter_98a}, 
and in a very recent work to the
$\gamma^{\ast}\gamma^{\ast}$-interaction~\cite{donnachie_98b}. 

In the color-dipole picture of high-energy scattering the cross section 
depends on the sizes of the scattered particles.
As a consequence, for instance the value of about 2/3 for the ratio of
$\pi p$ to $pp$ total cross sections can naturally be explained by 
the different radii of $\pi$ and $p$.
Though the SVM gives a prediction for the dipole-dipole cross section as
function of the dipole sizes, the resulting cross sections are 
energy-independent.
In particular, the $s^{0.08}$ dependence of soft high-energy 
hadron-hadron scattering~\cite{donnachie_92} does not follow from 
the SVM.
(According to a recent work~\cite{cudell_97} a behaviour like 
$s^{0.094}$ gives a better fit to hadron-hadron cross sections.)
Any energy-dependence of the scattering amplitude of the SVM has to be 
incorporated in a phenomenological way.

In order to explain the increase of $F_2$ at low $x$ and finite $Q^2$
the soft Pomeron behaviour of hadron-hadron scattering is not 
sufficient. 
Therefore, in our treatment we keep the energy-independent prediction 
of the SVM as soft Pomeron and add, similar to other approaches 
(see e.g.~\cite{donnachie_98,rueter_98b}), a hard Pomeron 
component.
For quite some time the BFKL-mechanism 
(exchange of a gluon-ladder)~\cite{fadin_75} has been considered 
as microscopic explanation of a hard Pomeron exchange.
However, due to the poor convergence of this perturbative 
approach~\cite{fadin_98}, the status of the BFKL-Pomeron is more than 
ever unclear.
Because of this situation, and the fact that we are mainly interested 
in the behaviour of the soft Pomeron, we start from a simple ansatz for 
the hard Pomeron as derived by L\'opez and 
Yndur\'ain~\cite{lopez_80} and obtained by the leading order DGLAP 
evolution~\cite{dokshitzer_77} of $F_2$.
We modify the solution of the DGLAP equation by multiplying a 
phenomenological factor, leading to a parametrization which can be 
applied in the limit $Q^2 \to 0$ without introducing a singularity 
in the cross section of photoproduction.
As a consequence of this modification our hard Pomeron component
is no longer, strictly speaking, a solution of the DGLAP equation 
in the nonperturbative region of low $Q^2$.

In the analysis we neglect a contribution from meson exchange, being
aware of the fact that at $W = 10 \, \rm{GeV}$ the trajectories of 
$a_2$ and $f_2$ give rise to an effect of about $10\%$.
Nevertheless, a consideration of the meson exchange introduces new
parameters but has only minor influence on the main results of our
investigation.
Even though we make a fit to experimental data, we emphasize that in the
present work we are not aiming at a fine-tuning of parameters.

\section{Soft Pomeron}
The structure function $F_2$ is given by the sum of the longitudinal
and transverse total $\gamma^{\ast}p$ cross section in the form
\begin{equation}
F_2 = \frac{Q^2}{4 \pi^{2} \alpha_{QED}}(\sigma_L + \sigma_T) \,.
\end{equation}
The relevant cross sections due to the soft Pomeron exchange have been 
calculated in Ref.~\cite{gousset_98} from the imaginary part of the 
forward amplitude for elastic $\gamma^{\ast}p$ scattering in the SVM. 
We obtain the result by summing over the flavours $f$ of the 
$q \bar{q}$-fluctuation of the virtual photon, 
\begin{equation} \label{sigma_svm}
\sigma_{L/T}^{SVM} = \sum_f \sigma_{f,L/T}^{SVM} = \sum_f e_f^2
 \int_0^1 dz \int_{r_{cut}}^{\infty} dr \, r \, {\cal I}_{f,L/T}(z,r) \,,
\end{equation}
where $e_f = \hat{e}_{f} e$ ($e$: elementary charge) denotes
the charge of the different quark flavours. 
We take into account the three light quarks $u,d,s$.
In (\ref{sigma_svm}) $z$ is the longitudinal momentum
fraction of the quark and $r$ the modul of the two-dimensional
vector $\vec{r} = r (\cos \vartheta, \sin \vartheta)$ between quark and
antiquark. 
The use of a lower bound $r_{cut}$ in the
$r$-integration differs from the treatment in~\cite{gousset_98} 
and will be discussed in more detail below. 
The functions ${\cal I}_{f,L/T}(z,r)$ read
\begin{eqnarray} \label{integrand_svm}
{\cal I}_{f,L}(z,r) & = & 
 \frac{N_c}{4 \pi^2} \, 4z^2(1-z)^2 \, Q^2
 K_0^2 (\varepsilon_f r) \, J_p(z,r) \,,
 \nonumber \\
{\cal I}_{f,T}(z,r) & = &
 \frac{N_c}{4 \pi^2} \left \{ [z^2+(1-z)^2]
 \varepsilon_{f}^{2} K_1^2 (\varepsilon_f r)
 + m_f^2 \, K_0^2 (\varepsilon_f r) \right \} J_p(z,r) \,, \;\, 
 \textrm{with}
 \\
\varepsilon_f^2 & = & z(1-z) \, Q^2 + m_f^2(Q_{eff}^2) \,.
 \vphantom{\frac{1}{1}} \nonumber
\end{eqnarray}
These quantities are obtained from the absolute square of the virtual 
photon light cone wave functions, which contain the modified Bessel 
functions $K_0$ and $K_1$.
$J_p(z,r)$ represents the soft Pomeron induced cross section for scattering 
of a $q\bar{q}$ color dipole of size $r$ off the proton target in the 
SVM~\cite{dosch_97}.
For a general dipole-proton cross section the expressions in 
(\ref{integrand_svm}) are identical to those given in 
Ref.~\cite{nikolaev_91}.
In our approach $J_{p}(z,r)$ can be written as
\begin{equation} \label{jp}
J_p(z,r) = 2 \int_{0}^{2\pi} d \vartheta \int_{0}^{\infty} 
 db \, b \int_{0}^{1} dz_{p}\int \frac{d^{2} \vec{r}_{p}}{4\pi} 
 \, |\psi_{p}(r_{p})|^{2} \,
 J(b, z, \vec{r}, z_{p}, \vec{r}_{p}) \,,  
\end{equation}
with $b$ denoting the impact parameter between the color dipoles of the
photon and proton.
For simplicity we consider the proton in the quark-diquark picture
and make use of a Gaussian wave function,
\begin{equation} \label{wf_proton}
 \psi_{p}(r_{p}) = \frac{\sqrt{2}}{S_{p}} 
 e^{- r_{p}^{2}/4 S_{p}^{2}} \,.
\end{equation}
The extension parameter $S_p$ in (\ref{wf_proton}) and the {\it rms} 
radius of the proton are related according to 
$S_p = 2 r_{p,rms} /\sqrt{3}$.

The quantity $J(b, z, \vec{r}, z_{p}, \vec{r}_{p})$ in Eq.~(\ref{jp}) 
is the interaction amplitude for the scattering of two color dipoles.
In the SVM, $J$ depends only very weakly on the momentum fractions
$z$ and $z_{p}$. 
For small dipole sizes $r$ and $r_p$ one can completely neglect this 
dependence in $J$, and hence also in $J_{p}$.
For small $r$, the dipole-proton cross section shows the typical
dipole-behaviour, $J_p(z,r) \propto r^2$, while for larger values of 
$r$ the cross section is no longer proportional to $r^2$. 
Around $1 \, \textrm{fm}$ for instance, one obtains a shape like
$r^{1.5}$ \cite{dosch_97}.

The SVM relates the dipole-dipole amplitude $J$ in the nonperturbative 
gluonic vacuum to the nonlocal gluon condensate.
For details about the computation of $J$ we refer the reader to the
literature (see e.g. \cite{dosch_94b,dosch_97}).
Here we only specify the field strength correlator.
Assuming that 
$\langle F_{\mu\nu}^{a}(z;w) F_{\rho \sigma}^{b}(0;w) \rangle$ 
does not depend crucially on the choice of the common reference point 
$w$, the most general form of the correlator reads
\begin{eqnarray} \label{correl}
g^{2} \langle F_{\mu\nu}^{a}(z;w) F_{\rho \sigma}^{b}(0;w) \rangle
 & = & \frac{\delta^{ab}}{N_{C}^{2} - 1} 
 \frac{1}{12} \langle g^{2}FF \rangle 
  \Big \{ \kappa (g_{\mu\rho} g_{\nu\sigma} 
           - g_{\mu\sigma} g_{\nu\rho}) D(z;a)
\\ 
 & & + \frac{1}{2} (1 - \kappa) \Bigl[ 
   \partial_{\mu} (z_{\rho} g_{\nu \sigma} - z_{\sigma} g_{\nu \rho})  
 + \partial_{\nu} (z_{\sigma} g_{\mu \rho} - z_{\rho} g_{\mu \sigma})
   \Bigr] D_{1} (z;a) \Big \} \,.
\nonumber
\end{eqnarray}
The first tensor structure of the correlator is of non-Abelian type
and leads to confinement, whereas the second term is an Abelian 
tensor.
The shape of the correlation functions $D(z;a)$ and $D_{1}(z;a)$
is governed by the correlation length $a$.

It is now obvious that the size of the dipole-proton cross section
$J_p(z,r)$ is given by the parameters of the field strength 
correlator $(\langle g^2 FF \rangle \,,\, a \,, \, \kappa)$ and the 
extension parameter $S_p$ of the proton. 
The quantity $\kappa$ is taken from a lattice 
simulation~\cite{digiacomo_92}, while the remaining three parameters
are fixed by the experimental values of the total $pp$ cross section
and the slope of the differential $pp$ cross section, both taken at
a {\it cm} energy of $20 \, \textrm{GeV}$, and in addition by the 
phenomenological $q \bar{q}$-string tension 
$\rho = 8 \, \kappa \, a^2 \langle g^2 FF \rangle / 81 \pi$.
To be explicit we adopt the values~\cite{dosch_97},
\begin{equation}
 \langle g^2 FF \rangle = 2.49 \, \textrm{GeV}^4 \,, \;
 a = 0.346 \, \textrm{fm} \,, \;
 \kappa = 0.74 \,, \;
 S_p = 0.74 \, \textrm{fm} \,.
\end{equation}

In Eq.~(\ref{integrand_svm}) $\varepsilon_f$ denotes the extension
parameter of the photon.
It depends on the quark flavour through the quark mass, and thus each 
flavour contributes in a different way to the integrands 
${\cal I}_{f,L/T}$.
A crucial quantity in ${\cal I}_{f,L/T}$ is the $Q^2$-dependent quark
mass.
For large values of $Q^2$, the hadronic component of the photon is
a free $q\bar{q}$ pair, while at lower $Q^2$ usually vector meson 
dominance (VMD) is used.
In our approach, the photon is represented by a $q \bar{q}$ fluctuation
over the whole region of $Q^2$.
This picture of the photon has been studied in detail in 
Ref.~\cite{gousset_98} and leads automatically to an effective quark 
mass interpolating between a constituent quark and a current quark.
Making use of quark-hadron duality the effective quark mass can be
derived by comparing the phenomenological photon polarization tensor,
which is obtained from VMD-poles and the perturbative continuum, with the 
polarization tensor we get in our description of the photon.
Since in the photon wave function $Q^2$ appears together with the factor 
$z(1-z)$, the quark mass has been investigated as function of 
$Q^2_{eff}=4z(1-z)\,Q^2$. 
The parametrization of the light quarks is~\cite{gousset_98} 
\begin{eqnarray} \label{mass_light}
m_{u/d}(Q^2_{eff}) & = &
 R \cdot 0.22 \, (1 - Q^2_{eff}/Q_{0,u/d}^2) \, \textrm{GeV} \,, \;
 \textrm{for} \; Q^2_{eff} \le Q_{0,u/d}^2 = 0.69 \, \rm{GeV}^2 \,,
 \nonumber\\ 
m_{u/d}(Q^2_{eff}) & = & 
 0 \,, \; \textrm{for} \; Q^2_{eff} \ge Q_{0,u/d}^2 \,,
\end{eqnarray}
while for the strange quark one gets
\begin{eqnarray} \label{mass_strange}
m_{s}(Q^2_{eff}) & = & 
 R \cdot [ 0.15 + 0.16 \, (1 - Q^2_{eff}/Q_{0,s}^2) ] \, \textrm{GeV} \,, \;
 \textrm{for} \; Q^2_{eff} \le Q_{0,s}^2 = 1.16 \, \textrm{GeV}^2 \,, 
 \nonumber\\
m_{s}(Q^2_{eff}) & = &
 R \cdot 0.15 \, \textrm{GeV} \,, \; \textrm{for} \; 
 Q_{eff}^2 \ge Q_{0,s}^2 \,,
\end{eqnarray}
with a parameter $R = 1$.
In previous works on inclusive scattering~\cite{gousset_98} and on
vector meson production~\cite{kulzinger_98} the cross sections induced
by real photons have always been too low by about $10-15 \%$.  
This drawback can be removed by lowering the quark masses.
Therefore, in our numerical calculation we take $R = 0.87$ which
gives us the best fit to the data.
The mass reduction of $13\%$ is probably within the error bars which 
the values of $m_f$ in (\ref{mass_light},\ref{mass_strange}) actually 
have. 

For the lower bound of the $r$-integration in (\ref{sigma_svm}) we 
choose $r_{cut} = a$.
Our hard Pomeron already describes the physics of small color
dipoles, even though we do not yet have a dipole-formula for the hard 
cross section.
To keep at small distances the SVM part of the cross section in addition 
to the hard part certainly leads to a double counting in this region.
Of course, our specific separation in soft and hard physics is to 
some extend arbitrary.
In particular, one could try to improve the final result by fitting the
value of $r_{cut}$, which introduces however a new parameter. 
Moreover, e.g. lattice data for the field strength correlator show a 
clear deviation from the specific correlator used in the SVM
at distances below the correlation length
(c.f. Ref.~\cite{meggiolaro_98}), where the deviation is due to 
a manifest perturbative contribution.
This means that in the correlator a transition between soft and hard 
contributions appears at distances of the order $0.3$--$0.4 \, \rm{fm}$.
The cut of the SVM contribution is similar to the procedure proposed by
Rueter~\cite{rueter_98b} previously.
Nevertheless, the low distance physics is described in a different 
way in both approaches.

\section{Hard Pomeron}
Also for the hard Pomeron in principle a dipole description with an 
improved photon wave function containing gluons in addition to 
the $q \bar q$ pair holds. 
The scattering of these gluons on the proton is by far not trivial, for
their transverse momenta can be small. 
A procedure has to be developed to separate soft and hard contributions. 
As working hypothesis the gluons with small light cone energies 
(i.e. finite light cone momenta and small transverse momenta) are 
already in the parametrization of the gluon field strength correlator
of the SVM. 
There  remain only gluons with large light cone energies 
(i.e. very small light cone momenta and large transverse momenta). 
These may be treated perturbatively.

To model the contribution of the hard Pomeron we consider the evolution 
of a power-behaved $F_2$ as derived by L\'opez and 
Yndur\'ain~\cite{lopez_80}.
Perturbative QCD implies that to leading order in the running coupling 
the singlet structure function is of the form
\begin{eqnarray} \label{f2_hard}
F_{2}^{\rm pert}(x,Q^2) & = & 
 B_{2} \, \alpha_s(Q^2)^{-d_{+}(1 + \lambda)} \, x^{-\lambda} \, ,
 \; \textrm{where} \vphantom{\frac{1}{1}}
 \\
d_{+}(1 + \lambda) & = &
 \frac{1}{\beta_0} \left(\frac{12}{\lambda}-11-\frac{2}{9}\right) \,,
 \; \textrm{and} 
 \nonumber  \\ 
\beta_0 & = & 11 - \frac{2}{3} N_f \,, \; N_f = 3 \,. \nonumber
\end {eqnarray}
In Eq.~(\ref{f2_hard}) $d_{+}$ denotes the leading eigenvalue of 
the anomalous dimension matrix of the quark-singlet and gluon 
evolution kernel. 
The formula for $d_{+}$ is valid for $\lambda$ close to zero.
The quantities $B_{2}$ and $\lambda$ are free parameters.
Eq.~(\ref{f2_hard}) can be applied only for small values of $x$
and is based on a singular gluon input.
We emphasize that (\ref{f2_hard}) is compatible with Regge theory, since
the intercept $(1 + \lambda)$ of the hard Pomeron does not depend on
$Q^2$.

While in Ref.~\cite{lopez_80} the expression (\ref{f2_hard}) was
used above $Q^2 = 3 \, \textrm{GeV}^2$, in a recent work Adel,
Barreiro and Yndur\'ain~\cite{adel_97} have proposed to analyse $F_2$ 
also for lower values of $Q^2$ on the basis of (\ref{f2_hard}).
However, in this case a phenomenological modification of (\ref{f2_hard})
is required in order to get a finite cross section for photoproduction.
One possible modification is given in Ref.~\cite{adel_97}.
The authors make use of a specific freezing of the strong coupling,
\begin{equation} \label{alpha_mod}
\alpha_s(Q^2) \to \frac{4 \pi}
 {\beta_0 \, \ln \, ((Q^2 + \Lambda^2)/\Lambda^2)} \,.
\end {equation}
Moreover, $d_{+}$ has to be replaced using the 
self-consistency equation
\begin{equation} \label{self}
 d_{+}(1 + \lambda) = 1 + \lambda \,.
\end{equation}
One can now show immediately that by means of
(\ref{alpha_mod},\ref{self}) the total cross section of 
photoproduction
\begin{equation}
 \sigma_{\gamma p} = \frac{4 \pi^2 \alpha_{QED}}{Q^2}\, 
 F_2 \Bigg |_{Q^2 = 0}
\end{equation}
is finite.
In~\cite{adel_97} Eq.~(\ref{self}) has been solved leading to 
$\lambda = 0.47$.
In particular, in the case of photoproduction this value seems to be
too large as will become obvious in the next section.
Therefore, we also apply Eq.~(\ref{self}), but in contrast 
to~\cite{adel_97} we keep $\lambda$ as a free parameter in our fit.

In addition to this parametrization, we consider an alternative ansatz
for the hard component $F_{2}^{\rm hard}$.
We multiply $F_{2}^{\rm pert}$ in Eq.~(\ref{f2_hard}) by the 
phenomenological factor $(Q^{2}/(Q^{2}+M^{2}))^{1 + \lambda}$ and 
freeze the strong coupling in a way different from the expression 
in (\ref{alpha_mod}).
According to that, the hard contribution reads  
\begin{eqnarray} \label{f2_hard_2}
F_{2}^{\rm hard}(x,Q^2) & = & 
 C_{2} \, \tilde{\alpha}_{s}(Q^2)^{-d_{+}(1 + \lambda)} 
 \left(\frac{Q^2}{Q^2 + M^2}\right)^{1+\lambda} x^{-\lambda} \,,
 \; \textrm{with} \\
\tilde{\alpha}_{s}(Q^2) & = &
 \frac{4 \pi}{\beta_0 \, \ln \, ((Q^2 + M^2) / \Lambda_{QCD}^2)} \,,
 \nonumber
\end{eqnarray}
where we apply a conventional $\Lambda_{QCD} = 0.25 \, \textrm{GeV}$.
To keep the number of free parameters as small as possible the same 
quantity $M$ serves as freezing mass in $\tilde{\alpha}_s$ and as 
parameter in the factor $Q^2/(Q^2+M^2)$.
Therefore, our ansatz for $F_{2}^{\rm hard}$ contains only three 
free parameters.
\\
Formula (\ref{f2_hard_2}) avoids a relation between the
freezing mass and $\Lambda_{QCD}$ in the strong coupling.
At large $Q^2$, QCD evolution (to leading order) is restored, 
and no use of the approximation (\ref{self}) has to be made.
Because of these reasons we consider the parametrization 
(\ref{f2_hard_2}) as the most natural phenomenological extension of
(\ref{f2_hard}) allowing us to interpolate between $Q^2 = 0$ and 
higher values of $Q^2$. 
Already in the past various authors 
(see e.g. \cite{badelek_92,donnachie_98}) have exploited terms of the 
type $Q^2/(Q^2+M^2)$ in order to reach the correct behaviour of $F_2$ at 
low $Q^2$. 

\section{Fitting Inclusive Photo- and Electroproduction}
The complete ansatz for the structure function reads
\begin{equation} \label{f2_total}
F_2 = F_{2}^{\rm soft} + F_{2}^{\rm hard} \,,
\end{equation}
where the soft part $F_{2}^{\rm soft}$ represents the
contribution of the SVM as discussed in Sec.~2.
Experimental data for both $\sigma_{\gamma p}$ and $F_2$ are fitted
through Eq.~(\ref{f2_total}).
In practice we fix $F_{2}^{\rm soft}$ from the SVM and fit
the two sets of parameters ($B_{2}, \lambda, \Lambda$ 
or $C_{2}, \lambda, M$) to the difference of the
data and the soft Pomeron contribution.
Since the evaluation of $F_{2}^{\rm soft}$ requires tedious multiple 
integrations, the only free parameter of the soft Pomeron 
(quantity $R$ in Eqs.~(\ref{mass_light},\ref{mass_strange})) is
not actually fitted but rather optimized on a discrete set of numbers
obtained in separate calculations.
 
For the data on $F_2$ we use the kinematical cuts 
$Q^2 \leq 6.5$ GeV$^2$, $x \leq 0.01$ and 
$W \geq 10 \, \textrm{GeV}$.
The limitation in $Q^2$ is mainly due to the fact that the soft Pomeron
part does not satisfy the DGLAP equation.
Our expression for the hard Pomeron requires the cut in $x$.
Since the scattering amplitude of the SVM is obtained from an eikonal
approximation the limitation in $W$ becomes mandatory.
The fit contains $150$ data points obtained at 
HERA~\cite{h1a_97,zeusa_97,h1b_96,zeusb_96}, 8 data points from
NMC~\cite{nmc_97} and 43 data points from E665~\cite{e665_96}.
Furthermore, 21 photoproduction data~\cite{caldwell_78,hera_photo} 
are included under the condition $W \geq 10 \, \textrm{GeV}$.
\begin{figure}[thb]
\center
\includegraphics[width=10cm,scale=1.0,angle=-90]{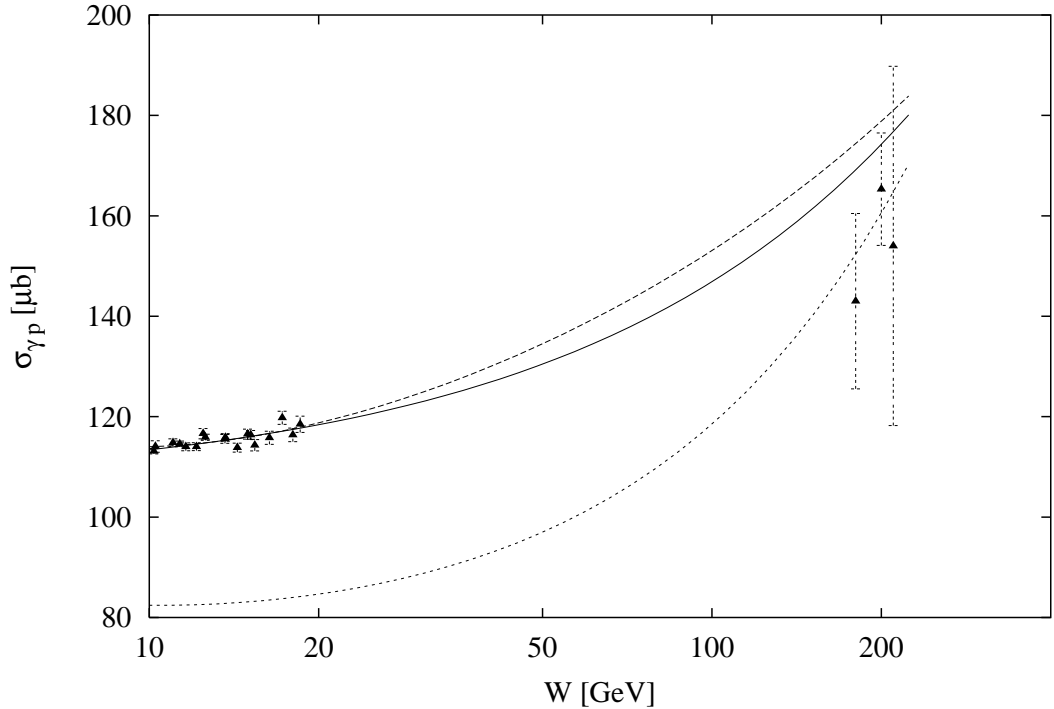}
\caption{\label{photo.fig}
 Total cross section for real photoproduction. Low-energy experimental
 points are from~\cite{caldwell_78}, high-energy points are 
 from~\cite{hera_photo}. 
 Our fit (full line) is compared to those performed by 
 Donnachie-Landshoff (dashed line)~\cite{donnachie_98} 
 and Adel-Barreiro-Yndur\'ain (dotted line)~\cite{adel_97}.}
\end{figure}

In the first parametrization a $\chi^2/\textrm{d.o.f.} = 1.00$, i.e. 
a good description of the experimental data is achieved.
In the calculation of $\chi^{2}$ the systematic and statistical errors
have been folded in quadrature.
The resulting values of the parameters are
\begin{eqnarray} \label{param_1}
B_{2} & = & 0.0268 \pm 6\% \,, 
 \nonumber \\
\lambda & = & 0.37 \pm 1\% \,,
 \nonumber \\
\Lambda & = & 1.12 \, \textrm{GeV} \pm 2\% \,. 
\end {eqnarray}
For the second parametrization the fit improves slightly.
We obtain $\chi^2/\textrm{d.o.f.} = 0.98$ with the parameters
\begin{eqnarray} \label{param_2}
C_{2} & = & 0.0025 \pm 3\% \,,
 \nonumber \\
\lambda & = & 0.37 \pm 1\% \,, 
 \nonumber \\
M & = & 1.02 \, \textrm{GeV} \pm 4\% \,.
\end {eqnarray} 

Obviously, the quality of the two fits is very similar.
The difference of both fit-functions becomes certainly more important as
soon as data at higher values of $Q^2$ are involved.
The errors of the parameters in (\ref{param_1},\ref{param_2}) are very 
small.
Our result for $\lambda$ is on the lower edge of the recent result
$(\lambda = 0.42)$ obtained by Donnachie and 
Landshoff~\cite{donnachie_98} 
and far below the value $\lambda = 0.47$ of Ref.~\cite{adel_97}. 
The numbers of the saturation scales $\Lambda = 1.12 \, \textrm{GeV}$ 
and $M = 1.02 \, \textrm{GeV}$ are quite similar to the typical scale
($1.2 - 1.5 \, \textrm{GeV}$) used in Ref.~\cite{badelek_92},
and may be related to the lowest hadronic state having a 
$q \bar q g$ or $q \bar q q \bar q$ structure.
These states could be considered as the entrance channel for the hard
Pomeron.
\begin{figure}[thb]
\center
\includegraphics[width=10cm,scale=1.0,angle=-90]{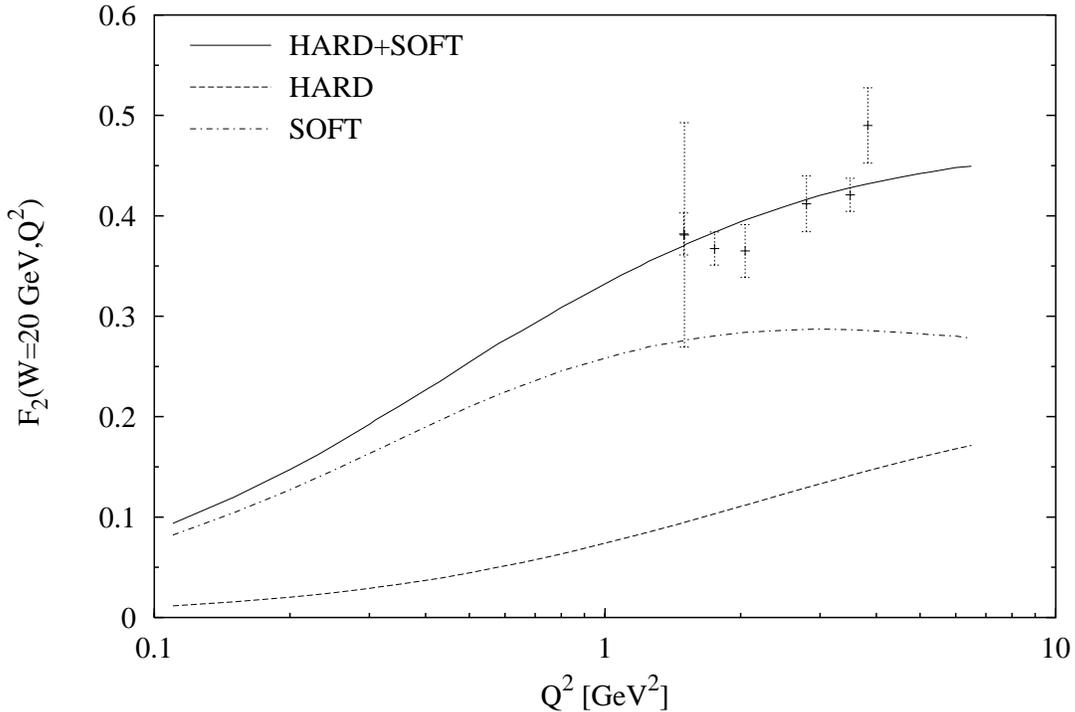}
\caption{\label{w20.fig}
 Structure function $F_2$ at fixed $W = 20 \, \textrm{GeV}$.
 The contributions of the soft and the hard Pomeron exchange are 
 shown separately.}
\end{figure}

In all numerical results we discuss in the following, our second
ansatz including the parameters of Eq.~(\ref{param_2}) enters.
We first consider the cross section for real photoproduction
(see Fig.~\ref{photo.fig}).
The soft Pomeron gives rise to the energy-independent contribution
$\sigma_{\gamma p}^{SVM} = 105.9 \, \mu{\rm b}$.
This number depends crucially on the value of the constituent quark 
mass, where a reduction of the quark mass increases the cross section.
In order to get in our two-component model a satisfying description 
of $\sigma_{\gamma p}$ for the whole energy range a reduction of the 
quark masses is unavoidable. 
The rise of $\sigma_{\gamma p}$ with increasing {\it cm} energy $W$
is completely given by the hard Pomeron.
This behaviour is different from the fit of Donnachie and 
Landshoff~\cite{donnachie_98}, where the hard Pomeron plays only a
subordinate role in real photoproduction and the shape of 
$\sigma_{\gamma p}$ is mainly determined by the $s^{0.08}$ dependence 
of the soft Pomeron contribution.

The parametrization of Adel, Barreiro and Yndur\'ain~\cite{adel_97}
is similar to our approach.
Contrary to us, these authors exploit for the soft Pomeron part in 
$F_2$ the simple VMD-inspired expression $Q^2/(Q^2 + \Lambda^2)$.
Their fit, which includes data on $F_2$ down to 
$Q^2 = 0.32 \, \textrm{GeV}^2$, can only describe the HERA data on 
$\sigma_{\gamma p}$ but underestimates the low energy data by
about $35 \%$.
One reason of this shortcoming is certainly the high value 
$\lambda = 0.47$ adopted in~\cite{adel_97}. 
\begin{figure}[thb]
\center
\includegraphics[width=10cm,scale=1.0,angle=-90]{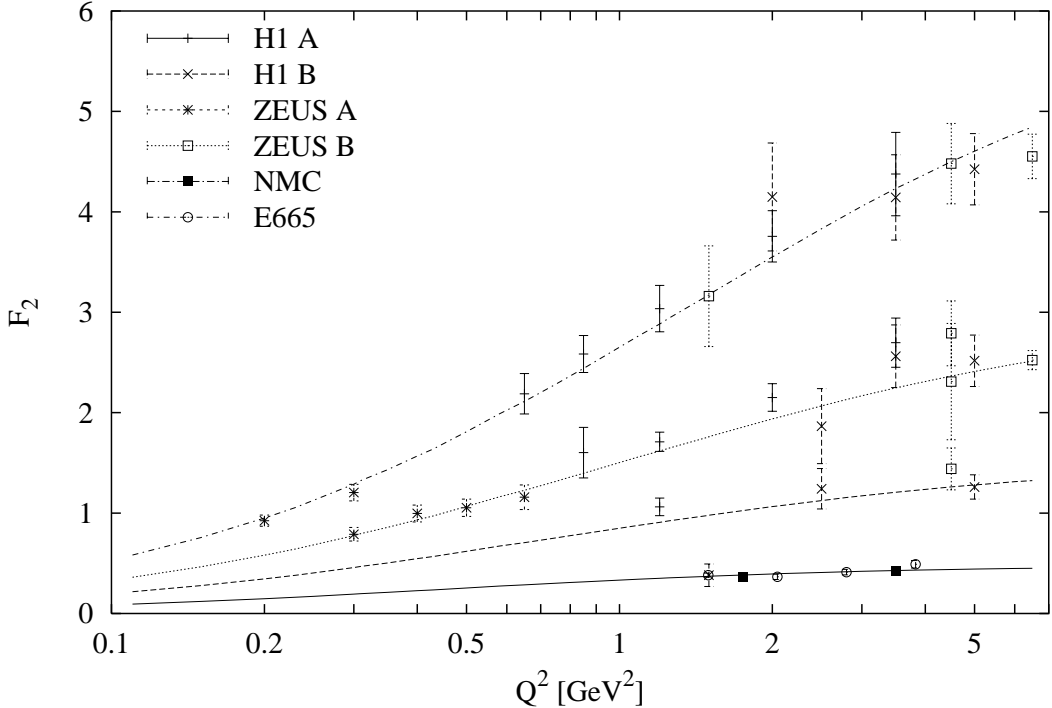}
\caption{\label{wfixed.fig}
 $F_2$ vs $Q^2$ at fixed {\it cm} energy $W$, 
 from bottom to top: $W = 20$ $(\times 1)$, 60 $(\times 2)$,
 100 $(\times 3)$, 200 $(\times 4) \, \textrm{GeV}$.
 The data points and curves are rescaled by the numbers in brackets. 
 Experimental points are: H1 A \cite{h1a_97}, H1 B \cite{h1b_96}, 
 ZEUS A \cite{zeusa_97}, ZEUS B \cite{zeusb_96}, NMC \cite{nmc_97} 
 and  E665 \cite{e665_96}.
 The {\it cm} energies for the experimental points lie within a 
 range of $\pm 5\%$ around the quoted numbers.}
\end{figure}
\begin{figure}[thb]
\center
\includegraphics[width=10cm,scale=1.0,angle=-90]{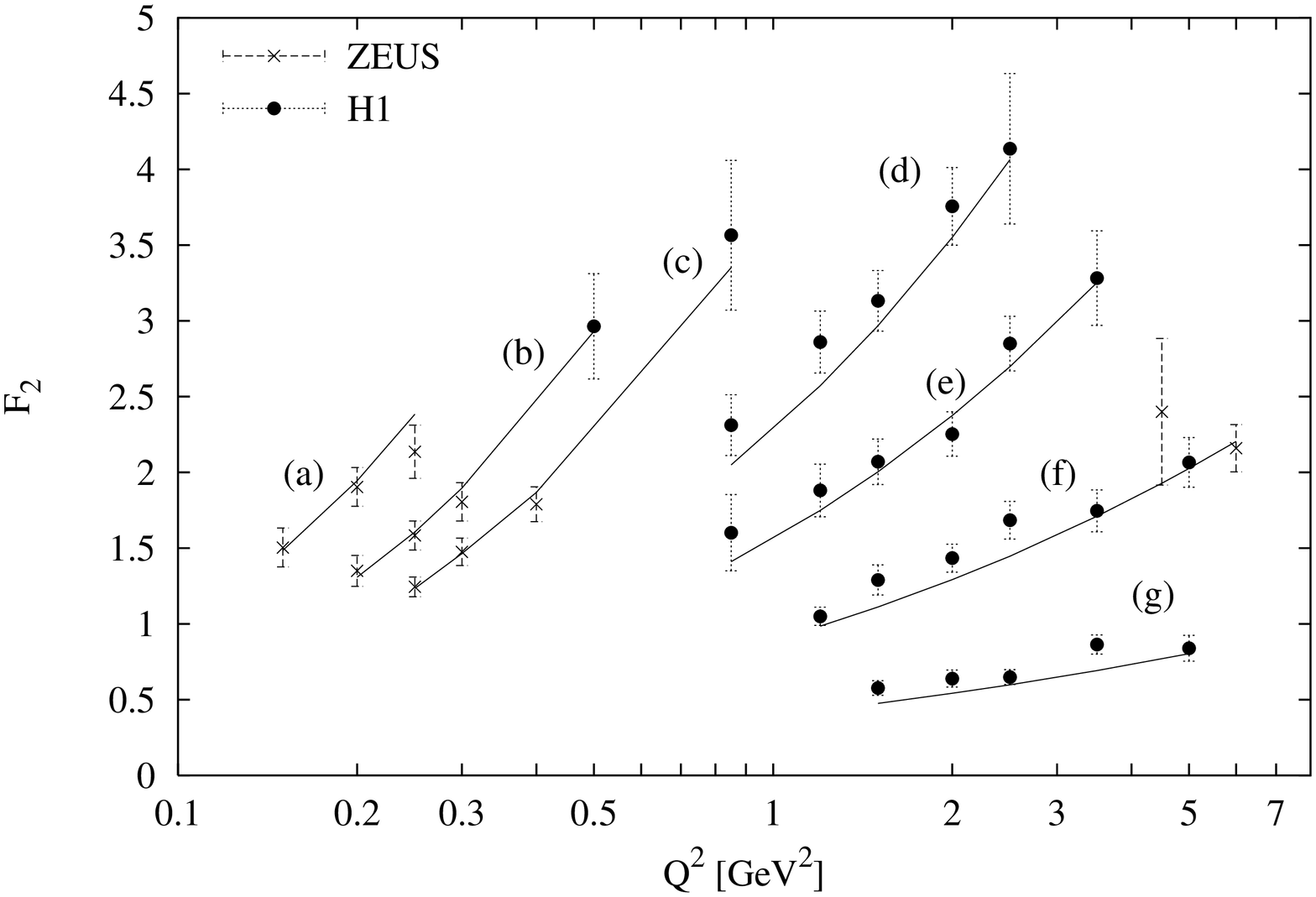}
\caption{\label{merino.fig}
 $F_2$ vs $Q^2$ at different values of $x$. 
 Experimental points 
 at (a), from left to right, $x=0.42\cdot 10^{-5}$, $x=0.44\cdot 10^{-5}$,
 $x=0.46\cdot 10^{-5}$ ($\times$ 8); 
 (b), from left to right, $x=0.85\cdot 10^{-5}$, $x=0.84\cdot 10^{-5}$,
 $x=0.83\cdot 10^{-5}$ and $x=0.86\cdot 10^{-5}$ ($\times$ 6);
 (c), from left to right, $x=0.13\cdot 10^{-4}$ and three points at
 $x=0.14\cdot 10^{-4}$ ($\times$ 5); 
 (d) $x=0.5\cdot 10^{-4}$ ($\times$ 4);
 (e) $x=0.8\cdot 10^{-4}$ ($\times$ 3);
 (f) $x=0.2\cdot 10^{-3}$ ($\times$ 2);
 (g) $x=0.5\cdot 10^{-3}$ ($\times$ 1).
 The data points and curves are rescaled by the numbers in brackets.}
\end{figure}

We now consider the results for $F_2$ by focusing first on the 
$Q^2$-dependence of the structure functions at fixed $W$.
Fig.~\ref{w20.fig}, showing $F_2$ at $W = 20 \,\textrm{GeV}$,
demonstrates a good agreement with the experimental data.
The hard and the soft contributions are shown separately.
At low $Q^2$, both $F_2^{\rm soft}$ and $F_2^{\rm hard}$ are increasing
with $Q^2$, where the soft part reaches a maximum around 
$2 - 3 \,\textrm{GeV}^2$.
Since at higher $Q^2$ the $q \bar{q}$ dipoles of the photon are
dominantly small, the decrease of $F_2^{\rm soft}$ in this
kinematical region is due to the lower bound $r_{cut}$ in the 
integration over the dipole sizes.
We emphasize that the shape of $F_2^{\rm soft}$ has a strong similarity 
with the purely empirical finding of Donnachie and 
Landshoff~\cite{donnachie_98}. 
It is interesting to note that the soft and the hard Pomeron exchanges
give sizable contributions for a relatively large range in $Q^2$.
The hard Pomeron leads already at $Q^2 = 1 \,\textrm{GeV}^2$ to an 
effect of about $25\%$, while on the other side the soft Pomeron
part is at $Q^2 = 6 \,\textrm{GeV}^2$ of the order $60\%$ and therefore
still very large. 

To give an impression of the $W$-dependence, $F_2(Q^2)$ is shown in 
Fig.~\ref{wfixed.fig} for different {\it cm} energies.
While $F_2^{\rm soft}$ is independent on $W$, the behaviour of the
hard part is given by 
$F_2^{\rm hard} \propto 1/x^{\lambda} \approx (W^2/Q^2)^{\lambda}$.
This leads to the fact that for high $W$ the hard part dominates
even at relatively low $Q^2$.
In comparison with the case of $W = 20 \,\textrm{GeV}$ we find that 
for $W = 200 \,\textrm{GeV}$ the hard Pomeron contributes about
$65\%$ at $Q^2 = 1 \,\textrm{GeV}^2$ 
and $80\%$ at $Q^2 = 6 \,\textrm{GeV}^2$.
\begin{figure}[thb]
\center
\includegraphics[width=13cm,scale=1.5,angle=-90]{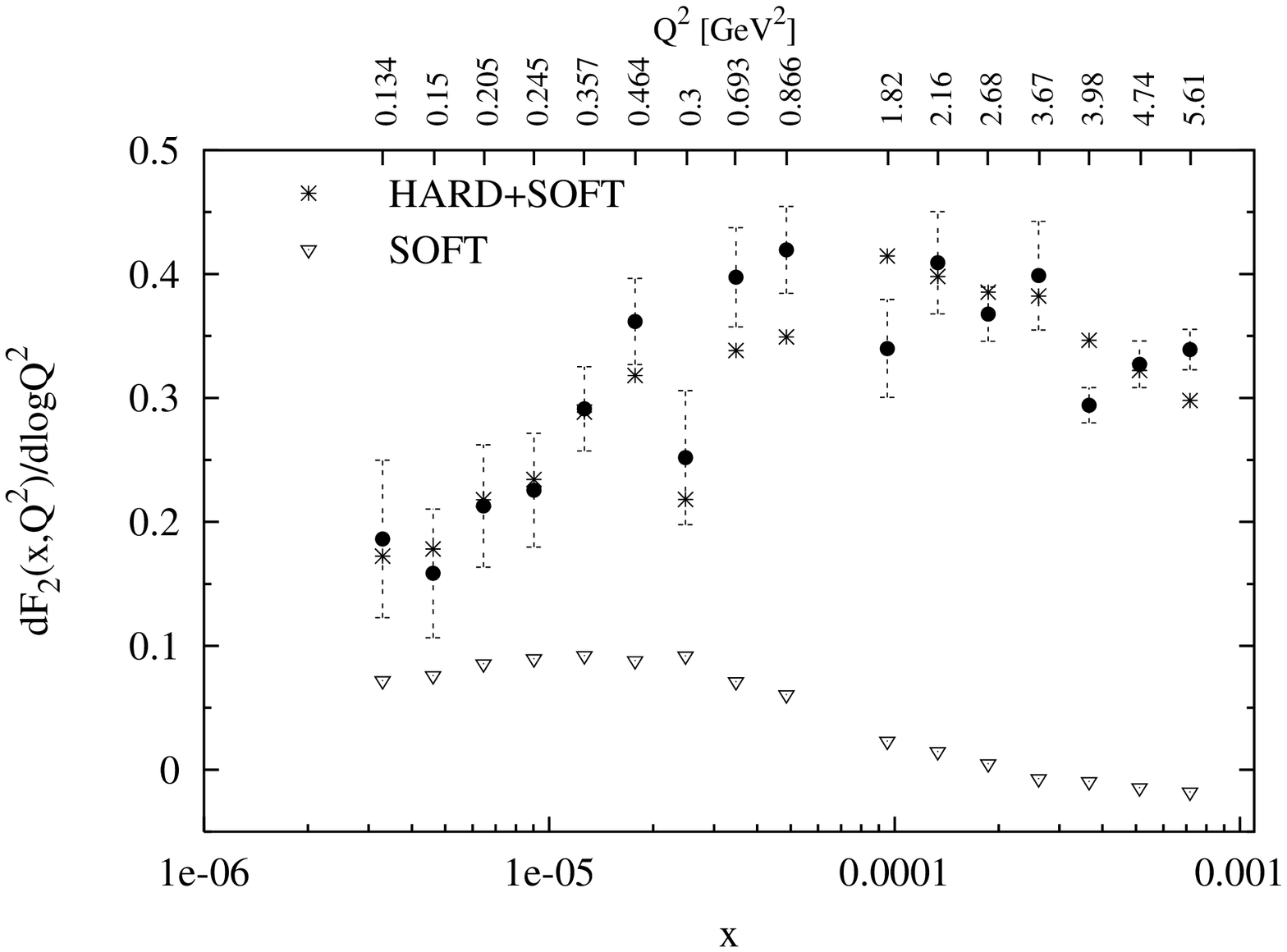}
\caption{\label{deriv.fig}
 Logarithmic derivative of $F_2$ vs $x$. 
 Data points are from \cite{caldwell_97}. 
 The soft contribution of the SVM is shown separately.}
\end{figure}

In Fig.~\ref{merino.fig} we plot $F_2(Q^2)$ for various values
of $x$ concentrating on the region of HERA kinematics.
This plot demonstrates that our model allows to fit the data for several
orders of magnitude in $x$.
To compare the relative contribution of the soft and the hard Pomeron
exchange we consider here for convenience only the experimental point 
with the lowest value in $x$ and $Q^2$ 
$(x = 0.42 \cdot 10^{-5}, \; Q^2 = 0.15)$.
In this case the soft part turns out to be of the order $55\%$.

Finally, we discuss the logarithmic slope 
$dF_2/d\,{\rm log}\,Q^2$ as shown in Fig.~\ref{deriv.fig}, where the
data points are taken from Ref.~\cite{caldwell_97}.
The experimental data in Fig.~\ref{deriv.fig} are usually considered 
as proof of a breakdown of the perturbative scaling violations as 
given by the DGLAP equation~\cite{glueck_95} at a certain $Q_0^2$.
However, as has been pointed out e.g. in Ref.~\cite{desgrolard_98b},
the value of $Q_0^2$ is strongly dependent on the specific selection 
of the experimental points.
Our two-component model explains the data on the derivative quite well. 
The contribution of the SVM is shown separately in Fig.~\ref{deriv.fig}.
At very low values of $Q^2$, the soft Pomeron gives rise to an effect
of about $50\%$.
This effect decreases with increasing $Q^2$ leading to a slightly
negative value above $2 - 3 \,\textrm{GeV}^2$.
The shape of the soft contribution just reflects the $Q^2$-dependence
of the SVM part shown in Fig.~\ref{w20.fig}.

\section{Longitudinal Structure Function}
\label{sec:5}
Without introducing any new parameter we are now able to compute the
longitudinal structure function $F_L$.
Making use of the relation
\begin{equation}
F_L = \frac{Q^{2}}{4\pi^{2} \alpha_{QED}} \, \sigma_{L}
\end{equation}
and Eq.~(\ref{sigma_svm}), the calculation of the SVM contribution
is straightforward.
The longitudinal cross section has to vanish in the limit 
$Q^{2} \to 0$.
In the SVM, where $\sigma_{L} \propto Q^{2}$ at low $Q^{2}$, this 
condition is automatically fulfilled. 
\\
To determine a hard component of $F_L$ we proceed as follows.
In perturbation theory the first nonvanishing contribution, arising from
the QCD compton process and boson-gluon fusion, is given
by~\cite{altarelli_78},
\begin{equation} \label{fl_pert}
F_{L}^{\rm pert}(x,Q^2) = \frac{\alpha_{S}(Q^2)}{2 \pi} x^{2}
 \int_{x}^{1} \frac{dy}{y^{3}} 
 \bigg[ \frac{8}{3} F_{2}^{\rm pert}(y,Q^2)
 + \, 4 \sum_{f} \hat{e}_{f}^{2} \, y g(y,Q^2) \Big( 1 - \frac{x}{y}\Big) 
 \bigg] \,.
\end{equation}
The gluon density $g$ in (\ref{fl_pert}) is related to the gluon 
structure function $F_G^{\rm pert}$ via
\begin{equation}
N_{f} F_{G}^{\rm pert}(x,Q^{2}) 
 = \sum_{f} \hat{e}_{f}^{2} \, x g(x,Q^{2}) \,.
\end{equation}
For low values of $x$, in Ref.~\cite{adel_97} both $F_{G}^{\rm pert}$ 
and $F_{2}^{\rm pert}$ have been determined on the same basis to 
leading order in the running coupling.
The two structure functions are related according to
\begin{eqnarray} \label{fg_pert}
F_{G}^{\rm pert}(x,Q^{2}) & = & 
 \frac{d_{+}(1 \! + \! \lambda) \! - \! D_{11}(1 \! + \!\lambda)}
 {D_{12}(1 + \lambda)} F_{2}^{\rm pert}(x,Q^{2})\,, \textrm{with}
 \nonumber \\
D_{11}(n) & = & 
 \frac{16}{3 \beta_0} \bigg[ \frac{1}{2 n(n+1)} + 
 \frac{3}{4} - n \sum_{k} \frac{1}{k(k+n)} \bigg] \,, 
 \nonumber \\
D_{12}(n) & = &
 \frac{2 N_{f}}{\beta_0} 
 \frac{n^{2} + n + 2}{n (n+1) (n+2)} \,. 
\end{eqnarray}
The quantities $D_{11}$ and $D_{12}$ are matrix elements of the 
anomalous dimension matrix, and $d_{+}$ is the eigenvalue as defined 
in (\ref{f2_hard}).
By means of the expressions in (\ref{f2_hard},\ref{fg_pert}) 
the structure function $F_{L}^{\rm pert}$ can easily be calculated.
To get a hard component of $F_{L}$ with an appropriate behaviour at 
low $Q^2$ we modify the perturbative result in the same spirit as we
have done in Eq.~(\ref{f2_hard_2}) for $F_{2}$.
This means, we multiply $F_{L}^{\rm pert}$ by the factor 
$(Q^{2}/(Q^{2}+M^{2}))^{2+\lambda}$, and moreover use the coupling
$\tilde{\alpha}_{s}$ in (\ref{f2_hard_2}).
Therefore, we finally obtain
\begin{eqnarray} \label{fl_hard}
F_{L}^{\rm hard}(x,Q^2) & = & \frac{C_{2}}{2\pi (2 + \lambda)} 
 \tilde{\alpha}_{s}(Q^{2})^{-d_{+}(1+\lambda) + 1} x^{- \lambda} 
 \\ 
& & \times \biggl[ \frac{8}{3} + \frac{4 N_{f}}{3 + \lambda}
  \frac{d_{+}(1 + \lambda) - D_{11}(1 + \lambda)}
 {D_{12}(1 + \lambda)} \biggr] 
 \biggl( \frac{Q^{2}}{Q^{2} + M^{2}} \biggr)^{2 + \lambda} \,.
 \nonumber
\end{eqnarray}
Obviously, by construction the behaviour of $F_{L}^{\rm hard}$ and
$F_{L}^{\rm soft}$ at low $Q^{2}$ coincides since both are proportional
to $Q^{4}$ near the real photon point. 

\begin{figure}[thb]
\center
\includegraphics[width=10cm,scale=1.0,angle=-90]{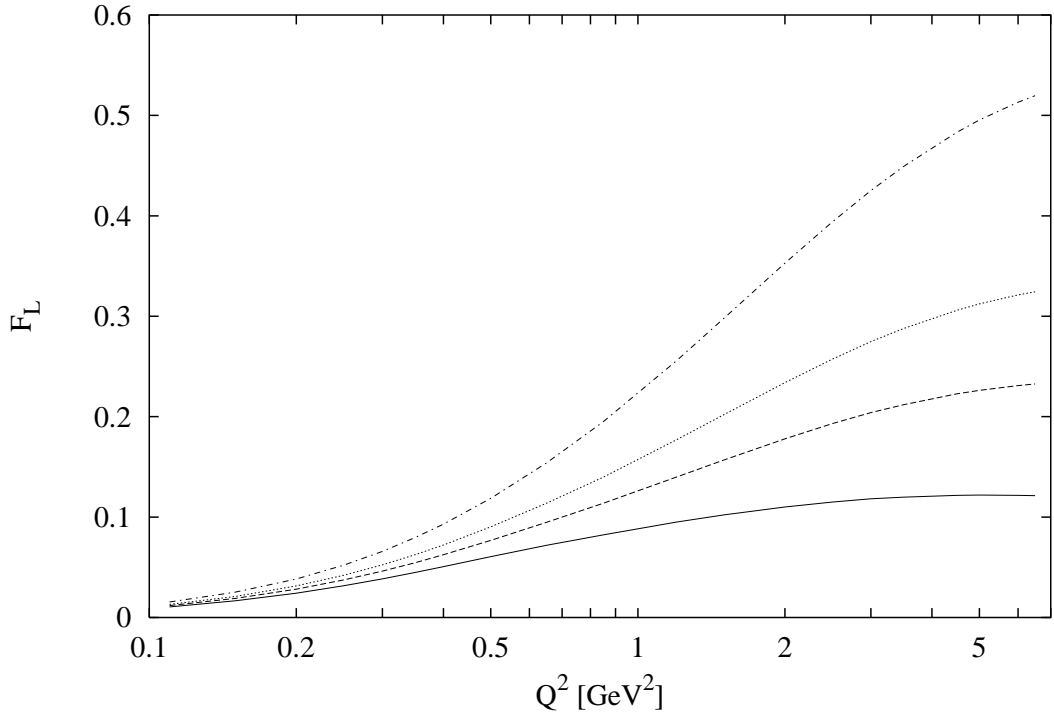}
\caption{\label{fl.fig}
 $F_L$ vs $Q^2$ at fixed {\it cm} energy $W$, 
 from bottom to top: $W = 20$, 60, 100 , $200 \, \textrm{GeV}$.}
\end{figure}

In Fig.~\ref{fl.fig} we plot $F_L$ as function of $Q^{2}$ for 
different values of the {\it cm} energy $W$.
At higher values of $Q^{2}$, the decrease of the energy-independent 
soft contribution $F_{L}^{\rm soft}$ is more marked than in the case of
$F_2$.
This behaviour arises since the $q \bar{q}$ dipole of a longitudinal 
photon, in average, is smaller than the hadronic fluctuation of a 
transverse photon. 
As a consequence, at $W = 200 \, \rm{GeV}$ and 
$Q^{2} = 6 \, \rm{GeV}^{2}$ the hard component exhausts almost 95\% of 
the total result. 

To compare our results with data we calculate the ratio
$R_{LT} = \sigma_{L}/\sigma_{T}$, which is the observable usually
measured in experiments.
In terms of the soft and hard components of $F_{L}$ and $F_{2}$ this
ratio can be written as
\begin{equation} \label{ratio}
R_{LT} = \frac{F_{L}^{\rm soft} + F_{L}^{\rm hard}}
 {(F_{2}^{\rm soft} - F_{L}^{\rm soft}) + 
  (F_{2}^{\rm hard} - F_{L}^{\rm hard})} \,.
\end{equation}
In the kinematical region of our fit there exist two data points from 
the NMC experiment~\cite{nmc_97}. 
As can be seen in Tab.~\ref{tab1}, our results agree fairly with these
data.
The agreement obviously adds confidence to our approach, even if it 
is clear that we are not able to really test the model with 
only two data points.

\begin{table}[h]
\begin{tabular}{cccc}
   $x \qquad$ & $Q^{2} \, [ \rm{GeV}^{2} ]$ & $R_{LT}^{exp}$ & 
   $\qquad R_{LT}$ 
     \\ \hline
   $0.0045 \qquad$ & $\qquad 1.38 \qquad$ & 
   $\qquad 0.537 \pm 0.129 \qquad$ & $\qquad 0.374$ \\ 
   $0.0080 \qquad$ & $\qquad 1.31 \qquad$ & 
   $\qquad 0.337 \pm 0.120 \qquad$ & $\qquad 0.347$ \\
  \hline
\end{tabular}
\caption{\label{tab1}
 Comparison of the ratio $R_{LT}$ in Eq.~(\ref{ratio}) with data 
 points from Ref.~\cite{nmc_97}.}
\end{table}

\section{Summary and Discussion}
We have presented a two-component model for inclusive 
$\gamma^{\ast}p$ scattering, which consists of a soft and a hard Pomeron
and is suitable in the region of low $x$ and low $Q^{2}$.
The four free parameters of the model have been adjusted to
the available data on the structure function $F_2$ of the proton 
(for $0.11 \le Q^2 \le 6.5 \, \textrm{GeV}^2$, $x \le 0.01$, 
$W \ge 10 \, \textrm{GeV}$) and on the total cross section of
real photoabsorption (for $W \ge 10 \, \textrm{GeV}$). 
The fit includes 222 data points and leads to the result 
$\chi^2/\textrm{d.o.f.} = 0.98$.

The soft Pomeron has been calculated from the Stochastic Vacuum Model,
which can be considered as an approximation of QCD in the infrared
region.
The SVM describes the complicated structure of the QCD vacuum
in terms of a nonlocal gluon condensate, where the variation of
the condensate in Minkowski space-time is governed by the correlation
length $a$.
In the framework of the SVM, diffractive scattering of two particles 
is equivalent to the scattering of two Wegner-Wilson-loops, leading 
automatically to cross sections in the color-dipole picture.
To fix the distribution of the loops in the transverse space, valence 
quark wave functions of the particles have to be introduced. 

The wave function of the photon has been determined in perturbation 
theory and accounts for the fluctuation of the $\gamma^{\ast}$ into a
$q \bar{q}$ state.
This description differs from VMD frequently used in the region of 
low $Q^2$.
A reasonable simultaneous description of $F_2$ and $\sigma_{\gamma p}$ 
for low and high $W$ by means of VMD is difficult, and requires in 
general further parameters.
VMD of the photon enters in our picture only through the determination 
of the quark masses by quark-hadron duality~\cite{gousset_98}, and
hence in a indirect way.

The soft Pomeron contains only one free parameter which regulates the
overall normalization of the $Q^2$-dependent quark masses in the photon
wave function.
Compared to previous work on $F_2$ at fixed
$W = 20 \,\textrm{GeV}$~\cite{gousset_98}, performed only with a soft
Pomeron, our fit favors a reduction of the quark masses by $13\%$.
Such a reduction improves also e.g. the cross section for 
photoproduction of $\rho$-mesons~\cite{kulzinger_98}.
The remaining (four) parameters of the soft Pomeron have been taken from 
other sources and left unchanged~\cite{dosch_97}.

The cross sections of the SVM are energy-independent, contrary to the 
$s^{0.08}$ behaviour of the soft Pomeron in hadron-hadron scattering.
To describe the data on $F_2$ obtained in fixed-target experiments and 
at HERA a hard Pomeron has to be considered in addition.
We have modeled a hard component by starting from the leading order 
QCD evolution of a power-behaved structure function $F_{2}$ 
($F_2 \propto x^{-\lambda}$)~\cite{lopez_80}.
Assuming a singular gluon input, the evolution does not produce a 
$Q^2$-dependence in the intercept, and hence the result is not in
conflict with Regge theory.
The result of the evolution has been multiplied by a simple 
phenomenological factor in order to obtain a finite cross section for real
photoproduction.
Our fit leads to $\lambda = 0.37$, which is close to a recently 
proposed value $(\lambda = 0.42)$ by Donnachie and 
Landshoff~\cite{donnachie_98}.

The parameters of the fit have been used to calculate also the
longitudinal structure function.
Like in the case of $F_{2}$ we have to modify the perturbative part of 
$F_L$, which serves as starting-point for the hard component, by a 
phenomenologal factor in order to enforce a vanishing 
$\sigma_{L}$ at $Q^{2} = 0$.
The numbers for the ratio $R_{LT} = \sigma_{L}/\sigma_{T}$ are in good
agreement with two data points from the NMC experiment. 
Up to now there exist no HERA data in the kinematical region of our fit.
However, recent activities at HERA will provide very soon final results
from a direct measurement of $F_L$ at low $Q^{2}$~\cite{eckstein_99}.

During the last time many people investigated $F_2$ at low $x$ and
especially at low $Q^2$ with different models.
The approaches comprise shadowing effects, Pomerons with a 
$Q^2$-dependent intercept, VMD calculations in combination with 
perturbative evolution and others 
(see e.g. Refs.~\cite{abramovicz_97,desgrolard_98a,martin_98,schildknecht_98,kaidalov_98,rostovtsev_98,golec_98,gotsman_98,buchmueller_98}).
Moreover, two-component Pomeron models have been applied to the 
$\gamma^{\ast}p$ interaction by various 
authors~\cite{adel_97,nikolaev_97,kerley_97,desgrolard_98b,donnachie_98,rueter_98b}.
With a soft and a hard Pomeron Donnachie and
Landshoff~\cite{donnachie_98} presented for a large kinematical
region a very good fit to $\sigma_{\gamma p}$ and $F_2$ using 10
parameters.
In this work not only the intercepts, but also the residues of both
Pomerons were fitted.
In contrast to this, the residue of our soft Pomeron has been fixed by 
the SVM and related to parameters of nonperturbative QCD.
In addition, at higher values of $Q^2$, the residue of the hard Pomeron 
follows the (leading order) evolution of QCD.

Our work strongly overlaps with the approach of Adel, Barreiro
and Yndur\'ain~\cite{adel_97}, since we are using essentially the same
expression for the hard Pomeron.
However, we differ in the way of performing the limit $Q^2 \to 0$ in the 
hard part and, in particular, in the ansatz of the soft Pomeron 
which is given by a single VMD-pole in~\cite{adel_97}.
The parametrization of Ref.~\cite{adel_97}, obtained by a fit to data on
$F_2$, fails in describing the data on $\sigma_{\gamma p}$ at low {\it cm} 
energies. 

The work of Rueter~\cite{rueter_98b}, where a good description of the
$\gamma^{\ast} p$ interaction was achieved, is also based on the 
SVM and therefore closest to ours.
As discussed in detail in Sec.~2, we cut the soft proton-dipole cross 
section below the correlation length $a = 0.346 \,\textrm{fm}$.
The interaction of small dipoles is taken into account by the hard
Pomeron.
A transition between soft and hard physics at distances of the order of
the correlation length is suggested by lattice calculations of the field
strength correlator~\cite{meggiolaro_98}.
In Ref.~\cite{rueter_98b} the treatment of the dipole-proton cross
section also changes for $r < a$.
Contrary to our approach, Rueter still makes use of the residue of the
soft Pomeron in the region of $0.16 - 0.35 \,\textrm{fm}$, but
multiplies for this kinematics the cross section by the energy-dependence
of a hard Pomeron (intercept 1.28).
Dipoles with an extension smaller than $0.16 \,\textrm{fm}$ are treated
by perturbative two-gluon exchange.

The extension of our two-component model to large $Q^2$ still has to be 
analysed.  
Moreover, one has to study the consequences in the case that our soft
contribution is multiplied by the energy-dependence of the soft Pomeron
of hadron scattering.
If the fit significantly improves we would interpret this result as a 
further hint that a soft Pomeron leading to a slight energy-increase 
is required not only in the interactions of hadrons  but also in 
$\gamma^{\ast}p$ interaction.
\bigskip

We thank H.G.~Dosch and M.~Rueter for critical discussions.
One of the authors (A.M.) thanks E.~Berger and G.~Kulzinger for
useful discussions concerning the model of the stochastic vacuum.
U.~D'Alesio was  funded through the European TMR Contract
No.~FMRX-CT96-0008: Hadronic Physics with High Energy Electromagnetic
Probes.
A.~Metz has been supported by BMBF.


\begin{thebibliography}{90}
\bibitem{h1a_97}
 H1: C. Adloff {\it et al.}, Nucl. Phys. {\bf B497}, 3 (1997).
\bibitem{zeusa_97}
 ZEUS: J. Breitweg {\it et al.}, Phys. Lett. {\bf B407}, 432 (1997).
\bibitem{gousset_98}
 H.G. Dosch, T. Gousset, and H.J. Pirner, Phys. Rev. {\bf D57}, 
 1666 (1998).
\bibitem{dosch_87}
 H.G. Dosch, Phys. Lett. {\bf B190}, 177 (1987).
\bibitem{dosch_88}
 H.G. Dosch and Yu.A. Simonov, Phys. Lett. {\bf B205}, 339 (1988).
\bibitem{dosch_94a}
 H.G. Dosch, Prog. Part. Nucl. Phys. {\bf 33}, 121 (1994).
\bibitem{dosch_94b}
 H.G. Dosch, E. Ferreira and A. Kraemer, Phys. Rev. {\bf D50}, 
 1992 (1994).
\bibitem{rueter_98b}
 M. Rueter, Eur. Phys. J. {\bf C7}, 233 (1999).
\bibitem{rueter_96}
 M. Rueter and H.G. Dosch, Phys. Lett. {\bf B380}, 177 (1996).
\bibitem{berger_98}
 E.R. Berger and O. Nachtmann, Eur. Phys. J. {\bf C7}, 459 (1999).
\bibitem{dosch_97}
 H.G. Dosch, T. Gousset, G. Kulzinger and H.J. Pirner, Phys. Rev.  
 {\bf D55}, 2602 (1997).
\bibitem{kulzinger_98}
 G. Kulzinger, H.G. Dosch and H.J. Pirner, 
 Eur. Phys. J. {\bf C7}, 73 (1999).
\bibitem{rueter_98a}
 M. Rueter, H.G. Dosch and O. Nachtmann, 
 Phys. Rev. {\bf D59}, 014018 (1999).
\bibitem{donnachie_98b}
 A. Donnachie, H.G. Dosch and M. Rueter, 
 Phys. Rev. {\bf D59}, 074011 (1999).
\bibitem{donnachie_92}
 A. Donnachie and P.V. Landshoff, Phys. Lett. {\bf B296}, 227 (1992).
\bibitem{cudell_97}
 J.R. Cudell, K. Kang and S.K. Kim, hep-ph/9712235 (1997).
\bibitem{donnachie_98}
 A. Donnachie and P.V. Landshoff, Phys. Lett {\bf B437}, 408 (1998).
\bibitem{fadin_75}
 V.S. Fadin, E.A. Kuraev and L.N. Lipatov, Phys. Lett. {\bf B60}, 50
 (1975); 
 Y.Y. Balitsky and L.N. Lipatov, Sov. J. Nucl. Phys. {\bf 28}, 822
 (1978).
\bibitem{fadin_98}
 V.S. Fadin and L.N. Lipatov, Phys. Lett. {\bf B429}, 127 (1998);
 M. Ciafaloni and G. Camici, Phys. Lett. {\bf B430}, 349 (1998);
 D.A. Ross, Phys. Lett. {\bf B431}, 161 (1998).
\bibitem{lopez_80}
 C. L\'opez and F.J. Yndur\'ain, Nucl. Phys. {\bf B171}, 231 (1980).
\bibitem{dokshitzer_77}
 Yu.L. Dokshitzer, Sov. Phys. JETP {\bf 73}, 1216 (1977);
 V.N. Gribov and L.N. Lipatov, Sov. J. Nucl. Phys. {\bf 15}, 78 (1972);
 G. Altarelli and G. Parisi, Nucl. Phys. {\bf B126}, 298 (1977).
\bibitem{nikolaev_91} 
 N.N. Nikolaev and B.G. Zakharov, Z. Phys. {\bf C49}, 607 (1991).
\bibitem{adel_97}
 K. Adel, F. Barreiro and F.J. Yndur\'ain, Nucl. Phys. {\bf B495}, 
 221 (1997).
\bibitem{donnachie_94} 
 A. Donnachie and P.V. Landshoff, Z. Phys. {\bf C61}, 139 (1994).
\bibitem{digiacomo_92}
 A. Di Giacomo and H. Panagopoulos, Phys. Lett. {\bf B285}, 133 (1992). 
\bibitem{meggiolaro_98}
 E. Meggiolaro, hep-ph/9807567 (1998).
\bibitem{badelek_92}
 B. Badelek and J. Kwieci\'nski, Phys. Lett. {\bf B295}, 263 (1992).
\bibitem{h1b_96}
 H1: S. Aid {\it et al.}, Nucl. Phys. {\bf B470}, 3 (1996).
\bibitem{zeusb_96}
 ZEUS: M. Derrick {\it et al.}, Z. Phys. {\bf C69}, 607 (1996);
 M. Derrick {\it et al.}, Z. Phys. {\bf C72} (1996) 399. 
\bibitem{nmc_97} 
 NMC: M. Arneodo {\it et al.}, Nucl. Phys. {\bf B483}, 3 (1997).
\bibitem{e665_96} 
 E665: M.R. Adams {\it et al.}, Phys. Rev. {\bf D54}, 3006 (1996).
\bibitem{caldwell_78}
 D.O. Caldwell {\it et al.}, Phys. Rev. Lett. {\bf40}, 1222 (1978).
\bibitem{hera_photo}
 H1: S. Aid {\it et al.}, Z. Phys. {\bf C69}, 27 (1995);
 ZEUS: M. Derrick {\it et al.}, Phys. Lett. {\bf B293}, 465 (1992);
 ZEUS: M. Derrick {\it et al.}, Z. Phys. {\bf C63}, 391 (1994).
\bibitem{caldwell_97}
 A. Caldwell, DESY Theory Workshop, DESY (1997).
\bibitem{glueck_95}
 M. Glueck, E. Reya and A. Vogt, Z. Phys. {\bf C67}, 433 (1995).
\bibitem{desgrolard_98b}
 P. Desgrolard, A. Lengyel and E. Martinov, 
 Eur. Phys. J. {\bf C7}, 655 (1999).
\bibitem{altarelli_78}
 G. Altarelli and G. Martinelli, Phys. Lett. {\bf B76}, 89 (1978).
\bibitem{eckstein_99}
 D. Eckstein, private communication (1999).
\bibitem{abramovicz_97}
 H. Abramowicz and A. Levy, hep-ph/9712415 (1997).
\bibitem{desgrolard_98a}
 P. Desgrolard, L. Jenkovszky and F. Paccanoni, 
 Eur. Phys. J. {\bf C7}, 263 (1999).
\bibitem{martin_98}
 A.D. Martin, M.G. Ryskin and A.M. Stasto, 
 Eur. Phys. J. {\bf C7}, 643 (1999).
\bibitem{schildknecht_98}
 D. Schildknecht, hep-ph/9806353 (1998).
\bibitem{kaidalov_98}
 A.B. Kaidalov and  C. Merino, hep-ph/9806367 (1998).
\bibitem{rostovtsev_98}
 A. Rostovtsev, M.G. Ryskin and R. Engel, 
 Phys. Rev. {\bf D59}, 014021 (1999).
\bibitem{golec_98}
 K. Golec-Biernat and M. Wuesthoff, 
 Phys. Rev. {\bf D59}, 014017 (1999).
\bibitem{gotsman_98}
 E. Gotsman, E. Levin, U. Maor and E. Naftali, 
 Nucl. Phys. {\bf B539}, 535 (1999).
\bibitem{buchmueller_98}
 W. Buchmueller, T. Gehrmann and A. Hebecker, 
 Nucl. Phys. {\bf B537}, 477 (1999).
\bibitem{nikolaev_97}
 N.N. Nikolaev, B.G. Zakharov and V.R. Zoller, JETP Lett. {\bf 66},
 138 (1997).
\bibitem{kerley_97}
 G. Kerley and G. Shaw, Phys. Rev. {\bf D56}, 7291 (1997).
\end{thebibliography}
\end{document}